\documentstyle[12pt]{article}

\begin{document}
\bibliographystyle{unsrt}
\textwidth 800pt
\large
\begin{center}
\underline{Movability of localized excitations in
nonlinear discrete systems - }
\\
\underline{a separatrix problem
}
\vspace{2cm}\\ \large S. \vspace{0.5cm}Flach and C. R. Willis \\
\normalsize
Department of Physics, Boston University,\\
Boston, Massachusetts 02215, \small flach@buphy.bu.edu \large
\vspace{1cm}
\\
\end{center}
\normalsize
ABSTRACT \\
We analyze the effect of internal degrees of freedom on
the movability properties of localized excitations
on nonlinear Hamiltonian lattices by means of properties
of a local phase space which is at least of dimension six.
We formulate generic properties of a movability separatrix
in this local phase space. We prove that due to the
presence of internal degrees of freedom of the localized
excitation it is generically impossible to define a
Peierls-Nabarro potential  in order to describe the motion of the
excitation through the lattice.
The results are verified analytically
and numerically for Fermi-Pasta-Ulam chains.
\vspace{0.5cm}
\newline
PACS number(s): 03.20.+i ; 63.20.Pw ; 63.20.Ry
\newline
{\sl Physical Review Letters}, in press.
\newline
Date: 11/19/93
\newpage
\normalsize
Recently localized breatherlike excitations were discovered to
exist in several different Hamiltonian lattices in one and two
dimensions \cite{st88},\cite{jbp90},\cite{bkp90},\cite{ff93},
\cite{dpw92},\cite{fw393}.
They are self-localized (no disorder) and appear
in nonlinear lattices - thus we name them nonlinear localized
excitations (NLEs). For certain systems it was possible to create moving
NLEs \cite{th91},\cite{ht92}.
Consequently the idea arose to describe their motion
in a Peierls-Nabarro-Potential (PNP) \cite{cp90},\cite{ckks93},
\cite{ysk93},\cite{kc93}
related to the PNP of kinks \cite{fn67},\cite{bsw88}.
Numerical simulations strongly support the existence
of a PNP-related phenomenon in Fermi-Pasta-Ulam systems \cite{sps92}
and Klein-Gordon systems \cite{dpw93}.
However as we show below it is generically impossible to
define a PNP for NLEs.

The NLE solutions are nontopological, i.e. no special structure
of the underlying many-particle potential is required. The only condition
is to have nonlinear terms in the potential. One can perform
stability analysis and show that if the NLE is localized enough
(in practice it will contain only a few particles which are involved
in the motion) then generically all Hamiltonian lattices will
exhibit families of stable time-periodic NLE solutions
\cite{fw393},\cite{fwo93},\cite{sa93}.
Hereafter we will call these stable periodic NLEs elliptic NLEs
to emphasize their
stability property (in a Poincare mapping they would appear
as stable elliptic fixed points \cite{fw393}).
One can
view the NLE as a solution of a reduced problem where only $M$
particles are involved in the motion, the rest of the lattice members
are held at their groundstate positions.
We showed that many frequency NLEs can be excited by perturbing the
elliptic NLEs and that thus NLE
solutions are motions on M-dimensional tori in the phase space
of the reduced problem and in the corresponding local subspace of
the phase space of the full system \cite{fw393},\cite{fwo93}.
Besides these stable NLE solutions unstable periodic
NLEs exist. Their feature is that certain local perturbations
destroy the unstable NLE or cause it to move
\cite{cp90},\cite{sps92},\cite{cku93}.
Hereafter we will call them
hyperbolic NLEs. If one calculates the energy density distribution
$e_l$
for the NLE solutions, one can define the position of the energy center
of the distribution by $X_E=\sum_l l\cdot e_l /(\sum_l e_l)$.
For a given
system the elliptic NLE solution yields $X_E=l_0$
(i.e. centered on a lattice site $l_0$) and the hyperbolic
NLE solution $X_E=l_1+0.5$
(i.e. centered between lattice sites $l_1$ and $l_1+1$) or vice versa
\cite{cp90},\cite{sps92},\cite{cku93}. Here $l,l_0,l_1$ denotes lattice
sites and the lattice spacing is 1. Both elliptic and
hyperbolic NLEs as well as certain stable subclasses of their perturbations
obey a symmetry during their whole evolution namely that the
evolution of the NLE part for $x < X_E$ is symmetric (or antisymmetric)
to the
evolution of the NLE part for $x > X_E$. This symmetry is just
the manifestation of Hamiltonian character of motion combined with
the discrete translational symmetry of the lattice.

The writing down of a certain PNP for the collective coordinate
which describes the motion of the NLE is conceptually equivalent to
the problem of a pendulum. The PNP-barrier $\Delta_{PN}$ is
then intimately connected with the energy that is required to
overcome the separatrix of the pendulum. This separatrix separates
oscillating pendulum solutions from rotating ones \cite{via73}. The PNP
frequency $\omega_{PN}$ is essentially the
pendulum frequency for infinitely small
amplitudes.

To describe a periodic elliptic NLE we need to introduce one
degree of freedom. We will work in the action-angle phase $(J,\theta)$
space and name this degree of freedom $J_1$. Its corresponding
frequency will be $\omega _1=\dot{\theta _1}=\partial H/\partial J_1$. Here $H$
denotes the full Hamiltonian of the lattice. We assume that
there exists a certain transformation between the original
variables (positions, momenta) and the actions and angles. This
does not imply integrability of the system as well as it does
not imply the inverse.
Since our NLE solutions
are regular solutions (at least on moderate time scales)
there is no need in introducing stochasticity
(cf.
\cite{fwo93} for details). Because of the symmetry of the
elliptic NLE the NLE will be stationary (nonmoving) for any value
of $J_1$ in the whole range of its existence. To excite
a moving NLE we have to excite an additional degree of freedom $J_3$.
Exciting $J_3$ we destroy the symmetry of the elliptic NLE.
But since
it is always possible to perturb the NLE conserving the symmetry,
we have to include an additional symmetryconserving degree of freedom $J_2$
into the
consideration. Thus we end up with the simplest generic case
of a Hamiltonian problem with three degrees of freedom:
\begin{equation}
H=H(J_1;J_2;J_3) \;\;\;, \;\;\; \omega_i=\dot{\theta_i}=
\frac{\partial H}{\partial J_i}, \;\;i=1,2,3 \;\;. \label{1}
\end{equation}
According to our notation $i=3$ labels the symmetrybreaking
degree of freedom. If it is excited strongly enough we expect
to hit a separatrix which separates stationary NLEs from
moving ones.
We will name this separatrix movability separatrix.
All three degrees of freedom are of local character,
especially they can be well defined in the reduced problem for the NLE.
Since we can consider the NLE excitation at (or between) any lattice
site(s) we thus study the local character of a movability separatrix which
is also defined for the infinite system. The movability separatrix for the
full system is just a periodic continuation of the local movability
separatrix.

Let us state the general condition for the
movability separatrix we are looking for. Since on the
movability separatrix
a trajectory will for inifinite times asymptotically reach a hyperbolic state
(which is nothing else then the hyperbolic NLE and its symmetric
perturbations) the corresponding frequency of the 3d degree
of freedom
\begin{equation}
\omega_3=\frac{\partial H}{\partial J_3}=f(J_1;J_2;J_3) \label{f}
\end{equation}
has to vanish on the movability separatrix i.e.,
\begin{equation}
f(J_1;J_2;J_3) = 0 \;\;\;, \label{2}
\end{equation}
which implies an equation for a surface in the three-dimensional
subspace of the actions $(J_1;J_2;J_3)$. We can always eliminate $J_2$
using the expression for the energy $E=H(J_1;J_2;J_3)$, so that
\ref{2} yields:
\begin{equation}
f(J_1;J_2;J_3)=\tilde{f}(E;J_1;J_3) = 0 \;\;\;.\label{3}
\end{equation}
{}From \ref{3} it follows that there exists a critical value
for $J_3$ on the movability separatrix:
\begin{equation}
J_3^s=g(J_1;J_2)=\tilde{g}(E;J_1) \;\;\;. \label{4}
\end{equation}
The only possibility of introducing the PNP would
be to use the relation between the potential of a pendulum and
its critical value for the action as well as the frequency of
small amplitude oscillations:
\begin{equation}
\omega_{PN}=f(J_1;J_2;J_3=0)=\tilde{f}(E;J_1;J_3=0) \;\;\;. \label{5}
\end{equation}
It is very important to note that if $f$ from \ref{f} or $\tilde{f}$
from \ref{3} depend on $(J_1;J_2)$ or $(E;J_1)$ respectively then
the same fact holds for $\omega_{PN}$ in \ref{5} as well as for
$J_3^s$ in \ref{4}.
As we immediately recognize a PNP would be different for
different $(E;J_1)$ because of the generic dependence of
the PNP parameters on the values of $E$ and $J_1$ in \ref{4},\ref{5}.
It is not only
that we would obtain different PNPs by varying the energy. Even for
a fixed energy different PNPs occur because of the dependence
of the right-hand sights in \ref{4},\ref{5} on $J_1$.

Let us discuss special nongeneric cases: i) The Hamiltonian
separates in the actions in the following way:
\begin{equation}
H(J_1;J_2;J_3)=H_{12}(J_1;J_2) + H_3(J_3) \label{6}
\end{equation}
Then according to its definition $\omega_3$ depends only
on $J_3$. Thus the value of $J_3^s$ becomes independent
on $(J_1;J_2)$ or $(E;J_1)$ and a unique PNP can be immediately
associated with the term $H_3(J_3)$ in \ref{6}.
ii) A more subtle nongeneric case appears if no separation
holds but the frequency $\omega_3$ is only a function of energy $E$.
In this case a PNP can be introduced which would depend on
the energy of the NLE.

Let us apply the results from above to a class of systems where
moving NLEs were detected \cite{th91},\cite{ht92},\cite{sps92}:
\begin{eqnarray}
H=\sum_l \left(\frac{1}{2}\dot{P}_l^2 + V(X_l -
X_{l-1})\right) \;\;, \label{7} \\
V(x)=\frac{1}{2}C x^2 + \frac{1}{4}x^4 \;\;. \label{8}
\end{eqnarray}
These systems belong to the class of Fermi-Pasta-Ulam systems \cite{jf92}.
$P_l$ and $X_l$ are momentum and position of the $l$-th particle
respectively. The parameter $C$ regulates the strength of the
quadratic terms. For $C \rightarrow \infty$, $E=const$ Eq.\ref{7}
becomes the well-known linear atomic chain, which is integrable and
has no NLE solutions. For $C \rightarrow 0$, $E=const$ Eq.\ref{7}
becomes a highly nonlinear nonintegrable atomic chain. All properties of
\ref{7} can be obtained by fixing the energy e.g. at $E=1$ and varying $C$.
All
solutions for other energies can be obtained by proper scaling
of the times, displacements and the parameter $C$: if \{$X_l(t;E=1;C)$\}
is a solution of \ref{7}, then \{$\tilde{X}_l(\tilde{t};\tilde{E};\tilde{C}$\}
is a solution for the energy $\tilde{E}=\lambda^{-4}$,
parameter $\tilde{C}=C/\lambda^2$ and
$\tilde{X}_l(\tilde{t})=\lambda^{-1}X(\lambda^{-1}t)$. Let us first
discuss the case $C=0$. Then an even simpler scaling holds - it
is enough to study the system at one given energy e.g. $E=1$ and
through the above described scaling all solutions for
other energies are obtained.
The elliptic NLE solution is the well known
even parity mode \cite{jbp90},\cite{sps92}.
It is centered between two particle sites
($X_E=l+0.5$) and four particles are essentially involved in the
motion. Its amplitude distribution can be qualitatively indicated
by ($\cdot$\hspace{1mm}$\cdot$\hspace{1mm}\small$\uparrow$\hspace{1mm}\LARGE$
\downarrow$\hspace{1mm}$\uparrow$\hspace{1mm}\small$
\downarrow$\large\hspace{1mm}$
\cdot$\hspace{1mm}$
\cdot$). \normalsize
More precisely the scaled absolute values of the  amplitudes
in decreasing order read: 1, 0.16579, 0.00048, ... .
No exact compacton structure is observed as it was wrongly
claimed in \cite{ysk93} because of a calculation error
in eq.13 of \cite{ysk93}.
The frequency of
the elliptic NLE for $E=1$ is $\omega_1(E=1)=1.760\pm0.0018$.
The hyperbolic
NLE solution is known as the odd parity mode \cite{sps92},\cite{cku93}.
It is
centered on a particle ($X_E=l$) and essentially three particles
are involved in the motion. Its amplitude distribution is roughly
($\cdot$\hspace{1mm}$\cdot$\hspace{1mm}\small$\downarrow$\LARGE\hspace{1mm}$
\uparrow$\hspace{1mm}\small$\downarrow$\hspace{1mm}\large$
\cdot$\hspace{1mm}$\cdot$). \normalsize
More precisely the scaled absolute values in decreasing order read:
1, 0.52304, 0.02305, ... . The frequency of the hyperbolic NLE is found to
be $\omega_{h}(E=1)=1.751 \pm 0.0018$.

Let us mention
an important property of \ref{7}. Besides the energy conservation law
this system conserves the total mechanical momentum:
$\sum_l P_l = const$.
It is sufficient to study the system in the center of mass frame,
so that the total momentum vanishes and the center of mass doesn't move.
All other cases can be obtained by a Galilean boost in \ref{7}.
Since the NLE solution is localized, the
total mechanical momentum outside the NLE is zero.
Thus it has to be zero inside
too, i.e. our NLE solutions have to obey mechanical momentum conservation,
at least approximately. The consequence is that the elliptic NLE (four
particles) is described by $4-1=3$ degrees of freedom. That is exactly
our simplest generic problem as described above.

The properties of the perturbed elliptic NLE can be studied with
Poincare mappings for symmetry-preserving perturbations, i.e. for
$J_3=0$. Then we can consider a reduced problem where the particles
outside the NLE are fixed at position zero. This fixed boundary doesn't
break the momentum conservation because of the antisymmetry of the
perturbed NLE. The result for the Poincare map is shown in Fig.1.
The point in the middle of the map corresponds to the elliptic
fixed point solution. All torus intersections inside the diamond-like
structured torus correspond to stable two-frequency NLEs in the
full system (1000 particles).
Every torus in Fig.1 corresponds to a certain triple
of $(J_1;J_2;J_3=0)$. The fixed point (elliptic NLE) is
defined by $(J_1;J_2=0;J_3=0)$.
Thus we first arrive at the unambiguous result that a perturbation
of the elliptic NLE preserving the symmetry leads to two-frequency
NLE solutions ($J_1;J_2;J_3=0$). This is similar to NLE properties
in Klein-Gordon lattices \cite{fw393},\cite{fwo93}.

Now we excite the third degree of freedom $J_3 \neq 0$ which destroys
the symmetry of the elliptic NLE.
We choose a path
in phase space where $P_1(t=0)=-P_0(t=0)$, $P_2(t=0)=-P_{-1}(t=0)=s$,
$X_1(t=0)=-X_{-1}(t=0)=a$ and all other displacements/momenta
are equal to zero at $t=0$.
The total energy is still $E=1$. We work with 1000 particles.
Here the elliptic NLE is chosen to
be centered between the lattice sites $l=0$ and $l=1$
respectively. The actions are some functions of the choosen
path: $J_1=J_1(E;s;a)$ and $J_3=J_3(E;s;a)$.
Especially we know that $J_3(E,s,a=0)=0$.
By increasing $a$ we measure the time dependence of the energy center
$X_E(t)$. The energy density is defined by
\begin{equation}
e_l= \frac{1}{2}P_l^2 + \frac{1}{2}(V(X_l-X_{l-1}) +
V(X_{l+1}-X_l))  \;\;\;. \label{el}
\end{equation}
Since $X_E(t)$ is independent of time for $a=0$,
we can hope that the energy center will essentially couple only
to $(J_3; \theta_3)$ so that we can measure the frequency
$\omega_3$. Indeed for $a \neq 0$ $X_E(t)$ oscillates
around its mean value of 0.5. There are modulations of this
oscillation with the frequency $\omega_1$,
but their amplitude is small and we clearly observe the frequency
$\omega_3=\tilde{f}(E;J_1(E;s;a);J_3(E;s;a))$.
For small values of $a$ ($< 10^{-4}$) the
value of $\omega_3$ becomes independent of $a$, thus we can
measure
$\tilde{f}(E;J_1(E;s;a=0);J_3=0)$ which is nothing else
than $\omega_{PN}$ (cf. \ref{5}).
Especially for the elliptic NLE solution we find $\omega_3=0.343
\pm 0.006$. For two other tori within the diamond like torus
in Fig.1 ($s=0.1$, $s=0.2$) we find $\omega_3=0.391 \pm 0.005$ and
$\omega_3=0.322 \pm 0.005$. Thus we find variations of
$\omega_3=\tilde{f}(E;J_1;J_3=0)$ for a
fixed value of the energy (by varying
$J_1$) of at least 21\%. Now let us increase $a$ for a given
value of $s$ and monitor the time evolution of $X_E(t)$. It
is shown in Fig.2. In agreement with our expectations we find
that with increasing $a$ the frequency $\omega_3$ decreases and
the amplitude of the oscillations of $X_E(t)$ increases. At
a threshold value of $a=a_s$ we clearly observe the crossing of
the movability separatrix - the NLE escapes from its mean position.

The properties of the movability
separatrix are easy constructed. Because
of the scaling property of the Hamiltonian for $C=0$ we
find all solutions at other energies by proper scaling. Since
the frequencies scale too, we immediately find that $\tilde{f}(
E;J_1;J_3=0)$ depends on the energy. Because we found strong (20\%)
variation of this frequency on the energy hypersurface (for constant
energy) the $J_1$-dependence is also significant.
Having $\tilde{f}(E;J_1;J_3=0)$ to be strongly dependent on $E$
and $J_1$  we find using \ref{f}-\ref{5} that the same holds
for the critical value of $J_3^s$ on the movability separatrix.
Thus we see
that our example is a generic case, and a PNP
can't be constructed.

If one considers $C\neq 0$ (here $C=0.3$) one rediscovers all
the above statements. There are only quantitative  changes -
the dependence of $\omega_3$ on $E$  and $J_1$ becomes weaker.
For large enough values of $C$ (fixing the total energy)
the frequencies $\omega_2$ and $\omega_3$ can become resonant
with the phonon band (which still doesn't prevent us from studying
the movability separatrix on short time scales). For too large values of
$C$ the frequency $\omega_1$ becomes resonant with the phonon band
and the whole NLE solution then quickly dissappears
\cite{bkpsss90},\cite{bks93},\cite{fw393},\cite{fwo93}.

Let us make some final comments. First our results demonstrate
a clear way of studying and characterizing the movability
properties of NLEs in terms of a movability separatrix in phase space.
Secondly we find that generically no simple PNP can be introduced.
The reason is the intimate connection
between the 'translational' degree of freedom ($J_3$) and
the 'internal' degrees of freedom ($J_1,J_2$) through
the Hamilton function.
Consequently the necessary energy supply to an elliptic NLE
in order to cross the movability separatrix at a certain
orbit can be positive, zero or negative depending on the
choosen orbit on the movability separatrix.
That is the reason why intuitive approaches to derive PNPs are
sometimes even selfcontradictory: in \cite{kc93} under
assumption of separability property (our equation \ref{6}) of a discrete
nonlinear Schr\"odinger equation a PNP is derived which is
energy dependent, but that implies the nonseparability of
the Hamiltonian.
The results in the present paper disprove
the conjecture in \cite{ysk93},
where it is predicted that for our equations \ref{7},\ref{8} and $C=0$
freely moving NLEs exist , i.e. no PNP (no separatrix) should exist.
Our Fig.2 shows that the separatrix exists.
A very interesting perturbation analysis was carried out in
\cite{ckks93},\cite{kc93}
for a weakly perturbed integrable Ablowitz-Ladik lattice. The authors
were able to show analytically that the NLE solution
is described by the evolution of
three internal degrees of freedom, so that
the movability separatrix can be analyzed analytically in their case.
Finally we mention the treatment of a discrete sine-Gordon breather
with a collective coordinate method \cite{bp91}. There it was shown
how to treat unambiguously a NLE with two degrees of freedom.
Already there it is clear that no unique PNP exists, i.e. the
amplitude of the PNP is a function of $E$ and the NLE amplitude.

This work was supported in part (S.F.) by the Deutsche Forschungsgemeinschaft
(Fl200/1-1).

\newpage
\begin{tabbing}
FIGURE CAPTIONS
\normalsize
\\
\\
\\
FIG.1 \hspace{1cm}\= Poincare intersection between the trajectory \\
\>of system \ref{7}\ref{8} for a reduced problem with fixed boundaries: \\
\> $C=0$, $E=1$, $P_1=-P_0$, $X_1=-X_0$, $P_2=-P_{-1}$, $X_2=-X_{-1}$, \\
\> all other lattice members are fixed at position zero. \\
\\
\\
\\
FIG.2 \> Time dependence of the center of energy of an NLE \\
\> for different assymetric perturbations $a$ (see text): \\
\> solid line - $a=0.02$; dotted line - $a=0.06$; dashed line - $a=0.1$; \\
\> long dashed line - $a=0.112$; dashed-dotted line - $a=0.113$.
\end{tabbing}

\newpage
\begin{thebibliography}{10}

\bibitem{st88}
A.~J. Sievers and S.~Takeno.
\newblock Phys. Rev. Lett.~61, 970 (1988).

\bibitem{jbp90}
J.~B. Page.
\newblock Phys. Rev.~B41, 7835 (1990).

\bibitem{bkp90}
V.~M. Burlakov, S.~A. Kisilev, and V.~N. Pyrkov.
\newblock Phys. Rev. B~42, 4921 (1990).

\bibitem{ff93}
F.~Fischer.
\newblock Ann. Physik~2, 296 (1993).

\bibitem{dpw92}
T.~Dauxois, M.~Peyrard, and C.~R. Willis.
\newblock Physica D~57, 267 (1992).

\bibitem{fw393}
S.~Flach and C.~R. Willis.
\newblock Phys. Lett. A~181, 232 (1993).

\bibitem{th91}
S.~Takeno and K.~Hori.
\newblock J. Phys. Soc. Japan~60, 947 (1991).

\bibitem{ht92}
K.~Hori and S.~Takeno.
\newblock J. Phys. Soc. Japan~61, 2186 (1992).

\bibitem{cp90}
D.~K. Campbell and M.~Peyrard. in:
\newblock {\em CHAOS - Soviet American Perspectives on Nonlinear Science, ed.
  by D. K. Campbell}.
\newblock American Institute of Physics New York, 1990.

\bibitem{ckks93}
C.~Claude, Yu.~S. Kivshar, O.~Kluth, and K.~H. Spatschek.
\newblock Phys. Rev.~B47, 14228 (1993).

\bibitem{ysk93}
Yu.~S. Kivshar.
\newblock Phys. Rev. E~48, R43 (1993).

\bibitem{kc93}
Yu.~S. Kivshar and D.~K. Campbell.
\newblock Phys. Rev. E~48, 3077 (1993).

\bibitem{fn67}
F.~Nabarro.
\newblock {\em Theory of Crystal Dislocations}.
\newblock Clarendon, Oxford, 1967.

\bibitem{bsw88}
R.~Boesch, P.~Stancioff, and C.~R. Willis.
\newblock Phys. Rev.~B38, 6713 (1988).

\bibitem{sps92}
K.~W. Sandusky, J.~B. Page, and K.~E. Schmidt.
\newblock Phys. Rev.~B46, 6161 (1992).

\bibitem{dpw93}
T.~Dauxois, M.~Peyrard, and C.~R. Willis.
\newblock unpublished~.

\bibitem{fwo93}
S.~Flach, C.~R. Willis, and E.~Olbrich.
\newblock Phys. Rev. E~49, 836 (1994).

\bibitem{sa93}
S.~Aubry.
\newblock Physica D~71, 196 (1994).

\bibitem{cku93}
O.~A. Chubykalo, A.~S. Kovalev, and O.~V. Usatenko.
\newblock Phys. Lett. A~178, 129 (1993).

\bibitem{via73}
V.~I. Arnol'd.
\newblock {\em Ordinary Differential Equations}.
\newblock MIT Press, Cambridge, 1973.

\bibitem{jf92}
J.~Ford.
\newblock Physics Reports~213, 271 (1992).

\bibitem{bkpsss90}
V.~M. Burlakov, S.~A. Kisilev, and V.~N. Pyrkov.
\newblock Solid State Comm.~74, 327 (1990).

\bibitem{bks93}
S.~R. Bickham, S.~A. Kisilev, and A.~J. Sievers.
\newblock Phys. Rev.~B47, 14206 (1993).

\bibitem{bp91}
R.~Boesch and M.~Peyrard.
\newblock Phys. Rev.~B43, 8491 (1991).

\end{thebibliography}
\end{document}